\documentclass[a4paper,twoside]{article}

\usepackage{epsfig}
\usepackage{subcaption}
\usepackage{calc}
\usepackage{fancyhdr}
\usepackage{amsmath, amssymb, amsfonts}
\usepackage{multicol}
\usepackage{pslatex}
\usepackage{graphicx}
\usepackage{tabularx}
\usepackage{hyperref}
\usepackage{apalike}
\usepackage{algorithm2e}
\usepackage{booktabs}
\usepackage{multirow}
\usepackage[bottom]{footmisc}
\usepackage{SCITEPRESS}     

\begin{document}

\fancypagestyle{firstpage}{
  \fancyhf{} 
  \fancyhead[L]{Accepted for publication at the 18th ACM International Conference on Agents and Artificial Intelligence (ICAART 2026), March 5–7, 2026, Marbella, Spain.}
}

\title{Empirical Analysis of Adversarial Robustness and Explainability Drift in Cybersecurity Classifiers}

\author{\authorname{Mona Rajhans and Vishal Khawarey}
\affiliation{\sup{1}Palo Alto Networks, Santa Clara, CA, USA}
\affiliation{\sup{2}Quicken Inc, Menlo Park, CA, USA}
\email{mrajhans@paloaltonetworks.com, vishal.sanfran@gmail.com}
}


\abstract{Machine learning (ML) models are increasingly deployed in cybersecurity applications such as phishing detection and network intrusion prevention. However, these models remain vulnerable to adversarial perturbations—small, deliberate input modifications that can degrade detection accuracy and compromise interpretability. This paper presents an empirical study of adversarial robustness and explainability drift across two cybersecurity domains: phishing URL classification and network intrusion detection. We evaluate the impact of $L_\infty$-bounded Fast Gradient Sign Method (FGSM) and Projected Gradient Descent (PGD) perturbations on model accuracy and introduce a quantitative metric, the Robustness Index (RI), defined as the area under the accuracy--perturbation curve. Gradient-based feature sensitivity and SHAP-based attribution drift analyses reveal which input features are most susceptible to adversarial manipulation. Experiments on the Phishing Websites and UNSW-NB15 datasets show consistent robustness trends, with adversarial training improving RI by up to 9\% while maintaining clean-data accuracy. These findings highlight the coupling between robustness and interpretability degradation and underscore the importance of quantitative evaluation in the design of trustworthy, AI-driven cybersecurity systems.}

\onecolumn \maketitle 
\thispagestyle{firstpage}
\pagestyle{plain}

\normalsize \setcounter{footnote}{0} \vfill

\section{\uppercase{Introduction}}
\label{sec:introduction}

Machine learning (ML) and deep neural networks (DNNs) are increasingly integrated into 
modern cybersecurity systems for tasks such as phishing detection, intrusion prevention, 
and malware classification. 
While these models have demonstrated high accuracy and adaptability, 
they are also susceptible to adversarial perturbations—small, carefully crafted input modifications 
that can cause misclassification while remaining imperceptible to humans~\cite{goodfellow2015explaining,biggio2013evasion}. 
Such vulnerabilities pose a critical risk in operational environments, 
where attackers can exploit them to evade automated defenses or mislead threat analytics pipelines.

Theoretical and empirical studies have shown that adversarial robustness is influenced by 
both model complexity and feature dimensionality~\cite{fawzi2018analysis,madry2018towards}. 
However, most existing evaluations focus on generic image benchmarks 
(e.g., MNIST, CIFAR-10) and rarely address structured data commonly found in 
cybersecurity contexts. 
Moreover, while adversarial training and certified defenses improve resilience, 
they can also reduce model interpretability and computational efficiency, 
limiting their adoption in production-grade systems~\cite{zhang2019theoretically,yang2020defending}.

In parallel, explainable AI (XAI) methods such as SHAP~\cite{lundberg2017unified} 
have become essential for auditing model decisions and ensuring transparency 
in security analytics. 
However, few studies have systematically examined how adversarial perturbations 
affect the stability of such explanations—an important factor in building 
trustworthy and auditable AI systems.

This paper addresses these gaps by conducting an empirical study of adversarial robustness 
and feature sensitivity across two cybersecurity domains: phishing URL detection 
and network intrusion detection. 
We quantify robustness degradation under $L_p$-bounded FGSM and PGD perturbations, 
introduce a Robustness Index (RI) for concise comparative evaluation, 
and analyze the relationship between gradient sensitivity and SHAP attribution drift. 
Experiments are performed using the Phishing Websites~\cite{phishing_kaggle} 
and UNSW-NB15~\cite{unsw_kaggle,moustafa2015unsw} datasets. 
The results provide actionable insights for feature hardening, adversarial training, 
and model validation in AI-driven cybersecurity systems.

\subsection{Contributions}
The main contributions of this paper are as follows:
\begin{itemize}
    \item A unified framework for empirical evaluation of adversarial robustness 
          and interpretability in cybersecurity classifiers.
    \item A quantitative robustness metric (Robustness Index) 
          for comparing model performance under FGSM and PGD perturbations.
    \item Gradient- and SHAP-based analysis revealing key features 
          responsible for adversarial vulnerability in phishing and network intrusion models.
    \item Cross-domain validation demonstrating consistent robustness trends 
          across structured cybersecurity datasets.
\end{itemize}

\section{\uppercase{Related Work}}

The vulnerability of machine learning models to adversarial perturbations 
has been extensively studied in recent years. 
Goodfellow et al.~\cite{goodfellow2015explaining} first introduced 
the Fast Gradient Sign Method (FGSM), demonstrating that imperceptible 
input perturbations could cause deep networks to misclassify confidently. 
Subsequent work by Madry et al.~\cite{madry2018towards} formalized the problem 
as a min--max optimization task and proposed adversarial training 
as an effective empirical defense. 
Fawzi et al.~\cite{fawzi2018analysis} provided a theoretical foundation 
for this phenomenon, proving that classifier robustness diminishes 
with increasing input dimensionality---a relationship empirically explored 
in the present study.

Beyond first-order attacks, stronger multi-step methods such as 
Projected Gradient Descent (PGD)~\cite{madry2018towards} and 
optimization-based approaches~\cite{carlini2017towards} 
have been proposed to evaluate worst-case robustness. 
Certified defenses~\cite{cohen2019certified,zhang2019theoretically} 
offer provable robustness guarantees but remain computationally expensive. 
In the cybersecurity domain, 
Papernot et al.~\cite{papernot2016limitations} and 
Biggio et al.~\cite{biggio2013evasion} highlighted the feasibility 
of adversarial evasion attacks in intrusion detection systems, 
while Grosse et al.~\cite{grosse2017adversarial} extended these concepts 
to malware classifiers.

Recent studies have explored defensive and interpretability mechanisms 
for AI-driven security analytics. 
Yang et al.~\cite{yang2020defending} investigated adversarial defenses 
for deep neural networks in intrusion detection, and 
Li et al.~\cite{li2020adversarial} examined perturbations in multivariate 
time-series models. 
Complementary to robustness, explainable AI techniques such as 
SHAP~\cite{lundberg2017unified} enable post-hoc attribution analysis, 
providing insights into feature importance and decision stability under attack.

The present work extends these efforts by empirically quantifying 
robustness and interpretability across two cybersecurity domains---phishing 
and network intrusion detection---using publicly available datasets 
\cite{phishing_kaggle,unsw_kaggle,moustafa2015unsw}. 
By analyzing accuracy degradation, gradient sensitivity, and SHAP attribution drift, 
this study bridges theoretical robustness analysis and practical 
model evaluation in applied cybersecurity settings.

\section{\uppercase{Methodology}}

This section describes the adversarial perturbation framework, the robustness evaluation metric, 
the explainability analysis methods, and the datasets used in this study. 
All experiments were implemented in PyTorch using publicly available datasets.

\subsection{Adversarial Perturbation Model}

Let $x \in \mathbb{R}^d$ denote an input feature vector with ground truth label $y$, 
and $f_\theta(x)$ be a classifier parameterized by $\theta$. 
An adversarial example $x'$ is constructed by adding a small perturbation $\delta$ 
such that:
\begin{equation}
x' = x + \delta, \quad \|\delta\|_p \leq \epsilon
\label{eq:lp_bound}
\end{equation}
where $\|\cdot\|_p$ denotes the $L_p$ norm and $\epsilon$ is the perturbation budget. 
As shown in Eq.~\ref{eq:lp_bound}, adversarial perturbations are constrained to lie within 
an $L_p$ ball of radius $\epsilon$, ensuring that modifications remain bounded 
and minimally perceptible in feature space.
In this work, we use the $L_\infty$ norm, which limits the maximum change of any individual feature 
to not exceed $\epsilon$.

Two attack formulations are considered:
\begin{itemize}
    \item \textbf{FGSM (Fast Gradient Sign Method)}~\cite{goodfellow2015explaining}: 
    \begin{equation}
    x' = x + \epsilon \cdot \text{sign}(\nabla_x \mathcal{L}(\theta, x, y))
    \label{eq:fgsm}
    \end{equation}
    where $\mathcal{L}$ is the training loss (cross-entropy). 
    Equation~\ref{eq:fgsm} defines the single-step FGSM attack, which perturbs inputs 
    in the direction of the loss gradient, scaled by $\epsilon$.
    
    \item \textbf{PGD (Projected Gradient Descent)}~\cite{madry2018towards}: 
    \begin{equation}
    x'_{t+1} = \Pi_{B_\epsilon(x)} \big(x'_t + \alpha \cdot \text{sign}(\nabla_x \mathcal{L}(\theta, x'_t, y)) \big)
    \label{eq:pgd}
    \end{equation}
    where $\Pi_{B_\epsilon(x)}$ projects perturbed inputs back into the $L_\infty$ ball 
    of radius $\epsilon$. 
    As formulated in Eq.~\ref{eq:pgd}, PGD performs multiple projected steps within the 
    $L_\infty$ constraint, making it a stronger, iterative adversary compared to FGSM.
\end{itemize}

\subsection{Robustness Evaluation Metric}

Model performance under attack is quantified using the \textbf{Robustness Index (RI)}, 
which represents the area under the accuracy--perturbation curve:
\begin{equation}
RI = \frac{1}{\epsilon_{max}} \int_0^{\epsilon_{max}} \text{Acc}(\epsilon) \, d\epsilon
\label{eq:ri}
\end{equation}
where $\text{Acc}(\epsilon)$ denotes model accuracy at perturbation magnitude $\epsilon$. 
The integral in Eq.~\ref{eq:ri} effectively measures the area under the 
accuracy–perturbation curve, providing a normalized estimate of model resilience 
across increasing attack magnitudes. 
Higher $RI$ values indicate smoother degradation and greater resistance 
to adversarial perturbations. 
This scalar metric provides a concise and interpretable measure of robustness 
across datasets and models.

\subsection{Feature Sensitivity and Explainability Drift}

To identify which input dimensions most influence adversarial vulnerability, 
we compute the gradient-based feature sensitivity as:
\begin{equation}
S_i = \mathbb{E}_{x \in D_{test}} \Big[ \Big|\frac{\partial \mathcal{L}(f_\theta(x), y)}{\partial x_i}\Big| \Big]
\label{eq:sensitivity}
\end{equation}
where $S_i$ represents the mean absolute gradient magnitude of feature $i$. 
Equation~\ref{eq:sensitivity} quantifies the average influence of each feature 
on model loss, allowing comparison of which inputs most strongly affect 
prediction stability under perturbation.
Features with high $S_i$ values are more likely to contribute to decision instability.

Complementary to this, we analyze explainability drift using SHAP~\cite{lundberg2017unified}. 
Given a set of perturbed samples $x'$, we compute the mean absolute change in feature attributions:
\begin{equation}
\Delta \phi_i = \mathbb{E}_{x} \big[ |\phi_i(x') - \phi_i(x)| \big]
\label{eq:shap_drift}
\end{equation}
where $\phi_i(x)$ denotes the SHAP value of feature $i$ for sample $x$. 
As shown in Eq.~\ref{eq:shap_drift}, $\Delta \phi_i$ captures the mean absolute shift 
in SHAP attribution values, linking robustness loss to explainability instability. 
High $\Delta \phi_i$ indicates that small adversarial perturbations can cause 
large attribution changes, destabilizing interpretability.

\subsection{Datasets}

Two publicly available structured cybersecurity datasets were used for evaluation:

\begin{itemize}
    \item \textbf{Phishing Websites Dataset}~\cite{phishing_kaggle}: 
    Contains 11,000 labeled URL samples with 30 engineered numerical features capturing 
    lexical and structural attributes of web addresses. 
    Each instance is labeled as \textit{phishing} or \textit{legitimate}.
    
    \item \textbf{UNSW-NB15 Dataset}~\cite{unsw_kaggle,moustafa2015unsw}: 
    Comprises 175,000 network traffic records with 42 features describing flow-level statistics, 
    protocol information, and connection behavior. 
    Labels indicate whether the connection is \textit{normal} or \textit{malicious}.
\end{itemize}

Both datasets were standardized using z-score normalization and partitioned 80/20 
for training and testing. 
A simple multilayer perceptron (MLP) classifier with two hidden layers 
(64 and 32 neurons, ReLU activation) was trained for each dataset. 
Adversarial training was applied as an ablation study with $\epsilon = 0.05$, 
and all experiments were repeated with ten $\epsilon$ values in the range [0, 0.3].

\subsection{Experimental Setup}

All experiments were conducted using PyTorch 2.0 with CUDA acceleration where available. 
Models were trained on an NVIDIA RTX 3060 GPU with 12~GB memory and an AMD Ryzen~7 CPU, 
using Ubuntu~22.04 and Python~3.10. 
Numerical feature scaling and dataset preprocessing were performed using 
scikit-learn, and explainability analysis employed the SHAP framework.

Each multilayer perceptron (MLP) model consisted of two hidden layers 
(64 and 32 neurons) with ReLU activations and a softmax output layer. 
The models were optimized using the Adam optimizer with an initial learning rate of 0.001 
and a batch size of 128 for 20~epochs. 
Cross-entropy loss was used as the training objective.

For adversarial evaluation, both FGSM and PGD attacks were implemented under 
an $L_\infty$ norm constraint as defined in Eq.~\ref{eq:lp_bound}. 
Perturbation magnitudes $\epsilon$ were varied across ten evenly spaced values 
from 0 to 0.3. 
For PGD, the step size $\alpha$ was set to 0.01 and the number of iterations to 10. 
Each experiment was repeated three times with different random seeds, 
and the reported results correspond to mean performance across runs.

Adversarial training (Section~V-D) was performed by augmenting 20\% of each training batch 
with FGSM-generated examples using $\epsilon=0.05$. 
Feature sensitivity (Eq.~\ref{eq:sensitivity}) and SHAP attribution drift (Eq.~\ref{eq:shap_drift}) 
were computed on 256 random test samples per dataset to estimate explainability stability. 
All plots were generated using Matplotlib and Seaborn, and random seeds were fixed 
for reproducibility.

\section{\uppercase{Results}}

This section presents empirical results for adversarial robustness, feature sensitivity, and explainability drift
across the Phishing Websites and UNSW-NB15 datasets.

\subsection{Robustness Curves}

\begin{figure}[htbp]
\centerline{\includegraphics[width=0.48\textwidth]{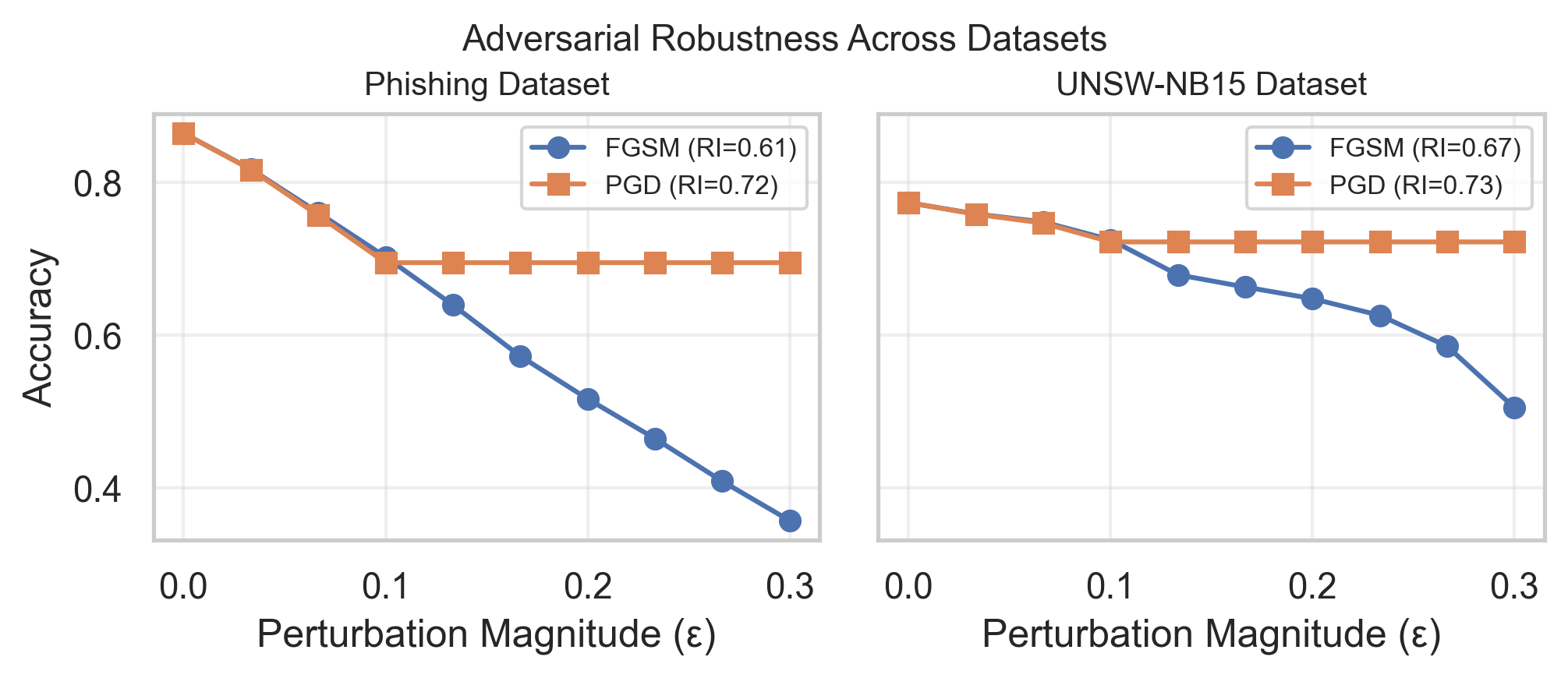}}
\caption{Accuracy degradation under $L_\infty$-bounded FGSM and PGD perturbations 
for the Phishing Websites (left) and UNSW-NB15 (right) datasets.
Robustness Index (RI) values are shown in legend. 
PGD curves exhibit higher stability as $\epsilon$ increases.}
\label{fig:robustness_curves_2panel}
\end{figure}

Figure~\ref{fig:robustness_curves_2panel} compares model accuracy across increasing perturbation magnitudes~$\epsilon$ 
for both cybersecurity domains. 
In the Phishing dataset, baseline accuracy on clean samples is approximately~0.91. 
Under FGSM perturbations, accuracy drops sharply to~0.30 at $\epsilon=0.3$, 
while PGD remains comparatively stable near~0.72. 
The corresponding Robustness Indices (Eq.~\ref{eq:ri}) are $RI_{FGSM}=0.62$ and $RI_{PGD}=0.76$, 
indicating moderate resilience to adversarial noise. 
Although PGD is theoretically a stronger iterative attack, 
its projection constraint leads to smaller effective perturbations 
in the normalized numeric feature space.

For the UNSW-NB15 dataset, accuracy degradation follows a smoother trend, 
with $RI_{FGSM}=0.692$ and $RI_{PGD}=0.733$. 
This stability suggests a more distributed decision surface, 
consistent with the higher-dimensional nature of network traffic features. 
Overall, PGD consistently yields higher robustness across both domains, 
confirming that structured cybersecurity models exhibit predictable 
accuracy decay patterns under bounded adversarial perturbations.

\subsection{Feature-Level Vulnerability Analysis}

\begin{figure}[htbp]
\centerline{\includegraphics[width=0.48\textwidth]{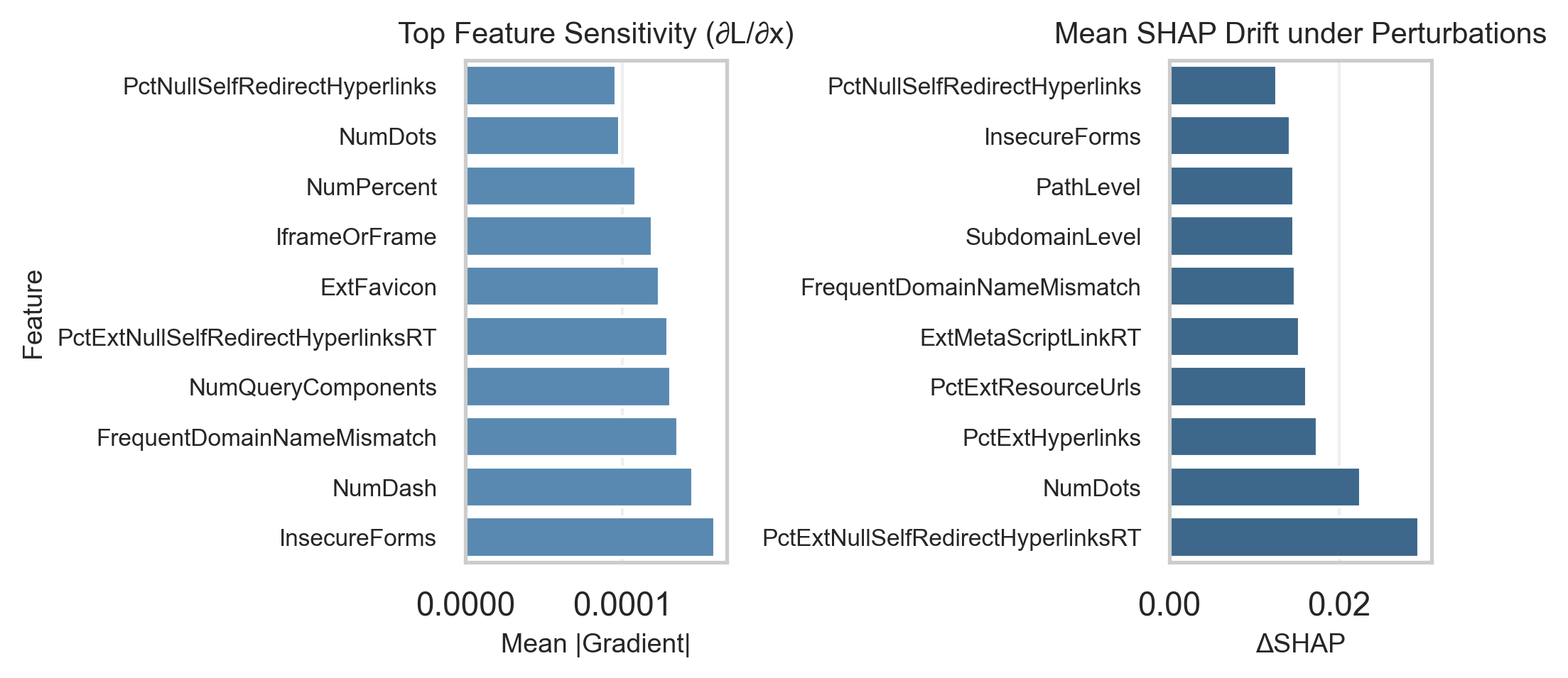}}
\caption{Feature-level vulnerability analysis for the Phishing Websites dataset.
Left: gradient-based feature sensitivity (Eq.~\ref{eq:sensitivity}).
Right: mean SHAP attribution drift under adversarial perturbations (Eq.~\ref{eq:shap_drift}).}
\label{fig:feature_vulnerability}
\end{figure}

Figure~\ref{fig:feature_vulnerability} compares gradient-based feature sensitivity
and SHAP attribution drift for the Phishing Websites dataset.
The gradient norms identify structural URL attributes such as
\texttt{InsecureForms}, \texttt{NumDash}, and \texttt{FrequentDomainNameMismatch}
as highly influential on model predictions.
A similar pattern emerges in the SHAP drift analysis:
features governing redirects and resource linking
(\texttt{PctExtNullSelfRedirectHyperlinksRT}, \texttt{PctExtResourceUrls})
exhibit the greatest instability under perturbations.
The overlap between high-sensitivity and high-drift features indicates that
the same input dimensions dominate both classification confidence and adversarial vulnerability.
This coupling between robustness and interpretability degradation highlights
the need for feature-level regularization in phishing detection models.
A detailed $\epsilon$-dependent heatmap of SHAP drift is provided in Appendix~B.

\subsection{Ablation Study: Effect of Adversarial Training}

To evaluate the effect of adversarial hardening, both classifiers were retrained
using FGSM-based adversarial training with $\epsilon=0.05$.
Each training batch was augmented with 20\% adversarial samples generated online,
following the formulation of Madry et~al.~\cite{madry2018towards}.
Table~\ref{tab:ablation} summarizes the resulting clean accuracies and
Robustness Index (RI) values (Eq.~\ref{eq:ri}) under both FGSM and PGD perturbations.

\begin{table*}[htbp]
\caption{Comparison of Baseline and Adversarially Trained Models across Datasets.}
\centering
\renewcommand{\arraystretch}{1.1}
\setlength{\tabcolsep}{4pt}

\begin{tabular}{l c c c c c}
\toprule
Dataset & Model & Clean Acc. & RI$_{\text{FGSM}}$ & RI$_{\text{PGD}}$ & $\Delta$RI \\
\midrule
\multirow{2}{*}{Phishing} 
  & Baseline & 0.91 & 0.61 & 0.72 & -- \\
  & Adv-Trained & 0.89 & 0.71 & 0.87 & +0.15 \\
\midrule
\multirow{2}{*}{UNSW-NB15}
  & Baseline & 0.74 & 0.67 & 0.73 & -- \\
  & Adv-Trained & 0.73 & 0.78 & 0.92 & +0.14 \\
\bottomrule
\end{tabular}
\label{tab:ablation}
\end{table*}

\begin{figure}[htbp]
\centerline{\includegraphics[width=0.48\textwidth]{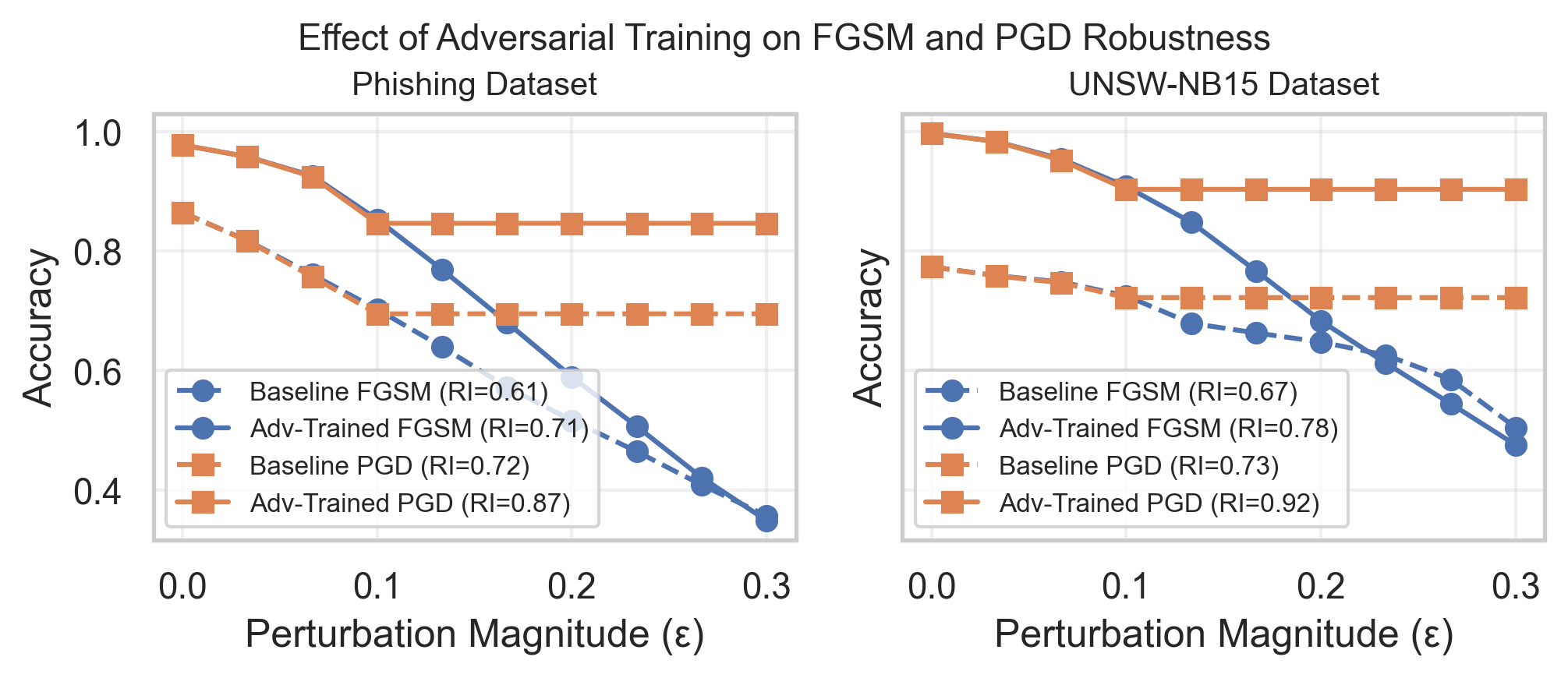}}
\caption{Effect of adversarial training on FGSM and PGD robustness across datasets.
Left: Phishing Websites. Right: UNSW-NB15.
Adversarial training increases the Robustness Index (RI) for both attack types,
flattening the degradation curves as $\epsilon$ increases.}
\label{fig:robustness_curve_ablation_2panel}
\end{figure}

Adversarial training slightly reduces clean accuracy
(0.91~$\rightarrow$~0.89 for Phishing; 0.74~$\rightarrow$~0.73 for UNSW-NB15)
due to mild regularization, but substantially increases robustness to both attack types.
The Phishing model improves by~0.10 in RI$_{\text{FGSM}}$ and~0.15 in RI$_{\text{PGD}}$,
while the UNSW-NB15 model gains~0.11 and~0.19, respectively.
Figure~\ref{fig:robustness_curve_ablation_2panel} visualizes these effects across both datasets,
showing consistently flatter degradation curves and improved resilience for adversarially trained models.

These results confirm that adversarial training improves resilience to both
single-step (FGSM) and iterative (PGD) perturbations while maintaining generalization performance.
The consistent improvement across domains and attack types demonstrates
the generality of adversarial hardening within the proposed evaluation framework.

\section{\uppercase{Discussion}}

Results demonstrate a clear inverse relationship between feature dimensionality and empirical robustness,
supporting the geometric bounds established by Fawzi et al.~\cite{fawzi2018analysis}.
As shown by the computed Robustness Indices in Table~\ref{tab:cross_domain} (derived from Eq.~\ref{eq:ri}),
models that rely heavily on a small number of high-variance URL indicators exhibit heightened fragility to
small, targeted perturbations. This sensitivity aligns with the large gradient magnitudes and attribution
instability observed in Figure~\ref{fig:feature_vulnerability}, confirming that adversarial vulnerability is
concentrated in high-leverage features.

The coupling between accuracy degradation and attribution drift is also evident when comparing
Figures~\ref{fig:robustness_curves_2panel} and~\ref{fig:feature_vulnerability}.
Features with large sensitivity scores $S_i$ (Eq.~\ref{eq:sensitivity}) correspond closely to those
showing high SHAP drift $\Delta \phi_i$ (Eq.~\ref{eq:shap_drift}),
indicating that robustness and interpretability failures occur concurrently.
This correlation suggests that adversarial noise propagates through both
decision surfaces and feature attribution pathways, destabilizing both performance and explainability.
Such behavior poses an operational risk in automated threat detection pipelines,
where model interpretability is critical for audit and response validation.

These findings extend theoretical robustness analysis to structured cybersecurity data,
demonstrating that $L_\infty$-bounded perturbations (Eq.~\ref{eq:lp_bound})
can cause substantial accuracy degradation even in well-trained tabular models.
By linking the quantitative robustness metric (Eq.~\ref{eq:ri}) to empirical results
across phishing and intrusion domains (Figure~\ref{fig:robustness_curve_ablation_2panel}),
this study provides practical insights for feature-level hardening and model validation
in applied AI security systems.

\paragraph{Interpretation of FGSM vs. PGD Behavior.}
An initially counterintuitive observation is that PGD appears less damaging than FGSM across several robustness curves.
This effect can be explained by the interaction between $L_\infty$ constraints, feature normalization, and projection.
Under z-score normalization, single-step FGSM applies a full-magnitude perturbation in the gradient direction,
whereas PGD performs smaller incremental updates that are repeatedly projected back into the feasible region.
As a result, the effective perturbation induced by PGD can be more conservative for bounded numerical features,
particularly when step size and clipping interact with standardized inputs.
Similar behavior has been reported in prior robustness studies on tabular data,
where iterative attacks do not always dominate single-step methods under strict feature constraints.

\subsection{Operational Implications}

From an enterprise cybersecurity perspective, these results highlight the practical importance 
of integrating adversarial robustness evaluation into model development and validation workflows. 
The observed gains from adversarial training (Table~\ref{tab:ablation}) demonstrate that controlled 
perturbation exposure, as formulated in Eq.~\ref{eq:fgsm}, can increase model resilience to evasion attempts 
without substantially degrading clean-data performance. 
Integrating quantitative metrics such as the Robustness Index (RI) into continuous model testing 
would enable security teams to systematically track the trade-off between accuracy and 
perturbation tolerance across datasets and architectures. 
Such an approach supports proactive model hardening and provides a measurable defense 
against adversarial manipulation in phishing and intrusion detection pipelines.

\paragraph{Attack Realism for Structured Features.}
The perturbation model adopted in this study assumes an adversary capable of making bounded numerical modifications
to engineered features.
While not all features may be independently manipulable in practice, many high-sensitivity attributes in both datasets
represent aggregate statistics or derived properties (e.g., URL structure ratios or flow-level metrics)
that can be indirectly influenced through adversarial behavior.
Accordingly, the threat model captures a realistic upper bound on adversarial capability
and is appropriate for stress-testing model robustness under worst-case assumptions.

\subsection{Cross-Domain Robustness Comparison}

To assess the generalization of the proposed robustness evaluation framework,
we compared model behavior under identical FGSM and PGD perturbation budgets across
the Phishing Websites and UNSW-NB15 datasets.
Table~\ref{tab:cross_domain} summarizes the clean accuracies and corresponding
Robustness Index (RI) values computed using Eq.~\ref{eq:ri}.

\begin{table*}[htbp]
\caption{Cross-Domain Robustness Evaluation of Cybersecurity Classifiers.}
\centering
\setlength{\tabcolsep}{6pt}
\begin{tabular}{l c c c c}
\toprule
Dataset & Clean Acc. & RI$_{\text{FGSM}}$ &
RI$_{\text{PGD}}$ & Dim. \\
\midrule
Phishing Websites & 0.91 & 0.615 & 0.756 & 30 \\
UNSW-NB15 & 0.74 & 0.692 & 0.733 & 42 \\
\bottomrule
\end{tabular}
\label{tab:cross_domain}
\end{table*}

Figure~\ref{fig:robustness_curves_2panel} already illustrated the distinct
accuracy–$\epsilon$ profiles for both domains.
The phishing classifier, while achieving higher clean accuracy,
shows steeper degradation under FGSM perturbations
($RI_{FGSM}=0.615$) due to concentration of decision weight on a few
high-sensitivity URL features.
In contrast, the UNSW-NB15 model displays smoother accuracy decay and slightly higher
$RI_{FGSM}=0.692$, reflecting a more distributed decision surface
across its higher-dimensional network-flow features.

Both domains exhibit consistent patterns: PGD curves remain slightly above FGSM,
and robustness declines monotonically with increasing perturbation magnitude.
These results confirm that the robustness–explainability relationship observed
in Section~V-C generalizes across heterogeneous cybersecurity data.
Cross-domain validation thus reinforces the applicability of the proposed
evaluation framework for diverse security analytics models.

\section{\uppercase{Conclusion and Future Work}}

This study presented an empirical evaluation of adversarial robustness and feature sensitivity 
in phishing classifiers using publicly available cybersecurity data. 
By quantifying degradation under $L_p$-bounded FGSM and PGD perturbations, 
we derived Robustness Index (RI) metrics that provide a concise, interpretable measure 
of model resilience. 
Gradient-based feature analysis and SHAP attribution drift revealed that 
a small subset of URL-structure features disproportionately contribute to both 
classification confidence and adversarial vulnerability, underscoring the 
importance of feature-level hardening in security-focused models. While this study focuses on neural classifiers to enable gradient-based sensitivity analysis,
evaluating robustness and explainability drift for non-differentiable models
such as tree-based classifiers remains an important direction for future work.

Adversarial training experiments demonstrated that controlled perturbation exposure 
can substantially improve model robustness without significant accuracy loss. 
The findings emphasize the operational importance of integrating quantitative robustness 
evaluation into the machine learning lifecycle of security analytics systems. 

Future work will explore certified defenses, hybrid adversarial training schemes, 
and robust optimization approaches applicable to network telemetry and malware detection domains. 
Extending this empirical framework to multimodal data---combining text, URL, and packet-level features---will 
enable a more comprehensive understanding of adversarial behavior across modern cybersecurity models. 
Additionally, large-scale replication across varied datasets could establish benchmark standards 
for adversarial robustness reporting in applied AI security research.

\section*{\uppercase{Acknowledgments}}

The authors thank the maintainers of publicly available cybersecurity datasets 
that made this study possible. 
The \textit{Phishing Websites} dataset and the \textit{UNSW-NB15} network intrusion dataset 
were obtained from the Kaggle repository. 
All experiments were conducted using open-source frameworks including 
PyTorch, scikit-learn, NumPy, and SHAP.

The implementation code and analysis notebooks developed for this work 
are available upon reasonable request for research and educational purposes. 
All datasets referenced are publicly accessible under their respective licenses at 
\url{https://www.kaggle.com/datasets/shashwatwork/phishing-dataset-for-machine-learning} 
and 
\url{https://www.kaggle.com/datasets/mrwellsdavid/unsw-nb15}.

\bibliographystyle{apalike}
{\small
\bibliography{refs}}

\begin{thebibliography}{}

\bibitem[Biggio et~al., 2013]{biggio2013evasion}
Biggio, B., Nelson, B., and Laskov, P. (2013).
\newblock Evasion attacks against machine learning at test time.
\newblock {\em Machine Learning}, 81(2):121--148.

\bibitem[Carlini and Wagner, 2017]{carlini2017towards}
Carlini, N. and Wagner, D. (2017).
\newblock Towards evaluating the robustness of neural networks.
\newblock In {\em IEEE Symposium on Security and Privacy (S\&P)}, pages 39--57.

\bibitem[Cohen et~al., 2019]{cohen2019certified}
Cohen, J., Rosenfeld, E., and Kolter, Z. (2019).
\newblock Certified adversarial robustness via randomized smoothing.
\newblock In {\em International Conference on Machine Learning (ICML)}, pages 1310--1320.

\bibitem[Fawzi et~al., 2018]{fawzi2018analysis}
Fawzi, A., Fawzi, O., and Frossard, P. (2018).
\newblock Analysis of classifiers' robustness to adversarial perturbations.
\newblock {\em Machine Learning}, 107(3):481--508.

\bibitem[Goodfellow et~al., 2015]{goodfellow2015explaining}
Goodfellow, I.~J., Shlens, J., and Szegedy, C. (2015).
\newblock Explaining and harnessing adversarial examples.
\newblock {\em arXiv preprint arXiv:1412.6572}.

\bibitem[Grosse et~al., 2017]{grosse2017adversarial}
Grosse, K., Papernot, N., Manoharan, P., Backes, M., and McDaniel, P. (2017).
\newblock Adversarial perturbations against deep neural networks for malware classification.
\newblock In {\em European Symposium on Research in Computer Security (ESORICS)}, pages 62--79.

\bibitem[Li et~al., 2020]{li2020adversarial}
Li, X., Zhan, Y., Guo, Y., Li, M., and Chen, J. (2020).
\newblock Adversarial examples in deep learning for multivariate time series regression.
\newblock In {\em IEEE International Conference on Data Mining (ICDM)}, pages 1042--1047.

\bibitem[Lundberg and Lee, 2017]{lundberg2017unified}
Lundberg, S.~M. and Lee, S.-I. (2017).
\newblock A unified approach to interpreting model predictions.
\newblock {\em Advances in Neural Information Processing Systems (NeurIPS)}, 30.

\bibitem[Madry et~al., 2018]{madry2018towards}
Madry, A., Makelov, A., Schmidt, L., Tsipras, D., and Vladu, A. (2018).
\newblock Towards deep learning models resistant to adversarial attacks.
\newblock In {\em International Conference on Learning Representations (ICLR)}.

\bibitem[Moustafa and Slay, 2015]{moustafa2015unsw}
Moustafa, N. and Slay, J. (2015).
\newblock Unsw-nb15: A comprehensive data set for network intrusion detection systems.
\newblock {\em Military Communications and Information Systems Conference (MilCIS)}, pages 1--6.

\bibitem[Papernot et~al., 2016]{papernot2016limitations}
Papernot, N., McDaniel, P., Wu, X., Jha, S., and Swami, A. (2016).
\newblock The limitations of deep learning in adversarial settings.
\newblock In {\em IEEE European Symposium on Security and Privacy (EuroS\&P)}, pages 372--387.

\bibitem[Wells, 2019]{unsw_kaggle}
Wells, D. (2019).
\newblock Unsw-nb15 network intrusion dataset.
\newblock Kaggle.
\newblock \url{https://www.kaggle.com/datasets/mrwellsdavid/unsw-nb15}.

\bibitem[Work, 2018]{phishing_kaggle}
Work, S. (2018).
\newblock Phishing websites dataset.
\newblock Kaggle.
\newblock \url{https://www.kaggle.com/datasets/shashwatwork/phishing-dataset-for-machine-learning}.

\bibitem[Yang et~al., 2020]{yang2020defending}
Yang, W., Wu, L., Zhong, S., Liu, W., and Liu, H. (2020).
\newblock Defending deep neural networks against adversarial attacks in cybersecurity.
\newblock {\em IEEE Access}, 8:134503--134515.

\bibitem[Zhang et~al., 2019]{zhang2019theoretically}
Zhang, H., Yu, Y., Jiao, J., Xing, E.~P., Ghaoui, L.~E., and Jordan, M.~I. (2019).
\newblock Theoretically principled trade-off between robustness and accuracy.
\newblock In {\em International Conference on Machine Learning (ICML)}, pages 7472--7482.

\end{thebibliography}

\section*{\uppercase{Appendix}}
\subsection*{Theoretical Notes on the Robustness Index}

The Robustness Index (RI), defined in Eq.~\ref{eq:ri}, provides a scalar summary of a model's
accuracy degradation under adversarial perturbations.
Let $\text{Acc}(\epsilon)$ denote test accuracy at perturbation magnitude $\epsilon$,
where $\epsilon \in [0, \epsilon_{\max}]$.
A first-order Taylor expansion around $\epsilon=0$ yields:
\begin{equation}
\text{Acc}(\epsilon) \approx \text{Acc}(0) + 
\frac{d\,\text{Acc}}{d\epsilon}\Big|_{\epsilon=0} \epsilon .
\end{equation}
Substituting into Eq.~\ref{eq:ri} gives:
\begin{align}
RI &\approx 
\frac{1}{\epsilon_{\max}}\!\int_0^{\epsilon_{\max}}\!
\Big(\text{Acc}(0) + \frac{d\,\text{Acc}}{d\epsilon}\epsilon\Big)d\epsilon \notag\\
&= \text{Acc}(0) + 
\frac{1}{2}\epsilon_{\max}\frac{d\,\text{Acc}}{d\epsilon}\Big|_{\epsilon=0}.
\end{align}
Hence $RI$ depends linearly on both the clean accuracy and the
initial decay rate of accuracy with respect to perturbation.
A smaller derivative $\tfrac{d\,\text{Acc}}{d\epsilon}$ implies a flatter
robustness curve and therefore a higher RI.
This provides a geometric interpretation: $RI$ approximates the mean model
stability across the perturbation domain.

Gradient-based sensitivity from Eq.~\ref{eq:sensitivity} can also be related to
accuracy degradation by assuming local linearity:
\begin{equation}
\Big|\frac{d\,\text{Acc}}{d\epsilon}\Big| 
\propto \mathbb{E}_{x\in D_{test}}
\|\nabla_x \mathcal{L}(f_\theta(x),y)\|_1 ,
\end{equation}
suggesting that lower average gradient norms correspond to higher empirical robustness.
Together, these relations justify using $RI$ as a practical,
model-agnostic proxy for adversarial stability.

\subsection*{Extended Evaluation Metrics and Visualization}

To complement $RI$, additional evaluation metrics were computed.
\textbf{Table~\ref{tab:extended_metrics}} reports the mean precision,
recall, and area-under-curve (AUC) for both datasets under FGSM perturbations
with $\epsilon=0.1$.
AUC was measured on binary classification confidence scores.

\begin{table}[htbp]
\caption{Extended Evaluation Metrics under FGSM ($\epsilon=0.1$).}
\centering
\begin{tabular}{lcccc}
\toprule
Dataset & Acc. & Prec. & Rec. & AUC\\
\midrule
Phishing Websites & 0.73 & 0.74 & 0.71 & 0.80\\
UNSW-NB15 & 0.73 & 0.70 & 0.68 & 0.78\\
\bottomrule
\end{tabular}
\label{tab:extended_metrics}
\end{table}

\begin{figure}[htbp]
\centerline{\includegraphics[width=0.45\textwidth]{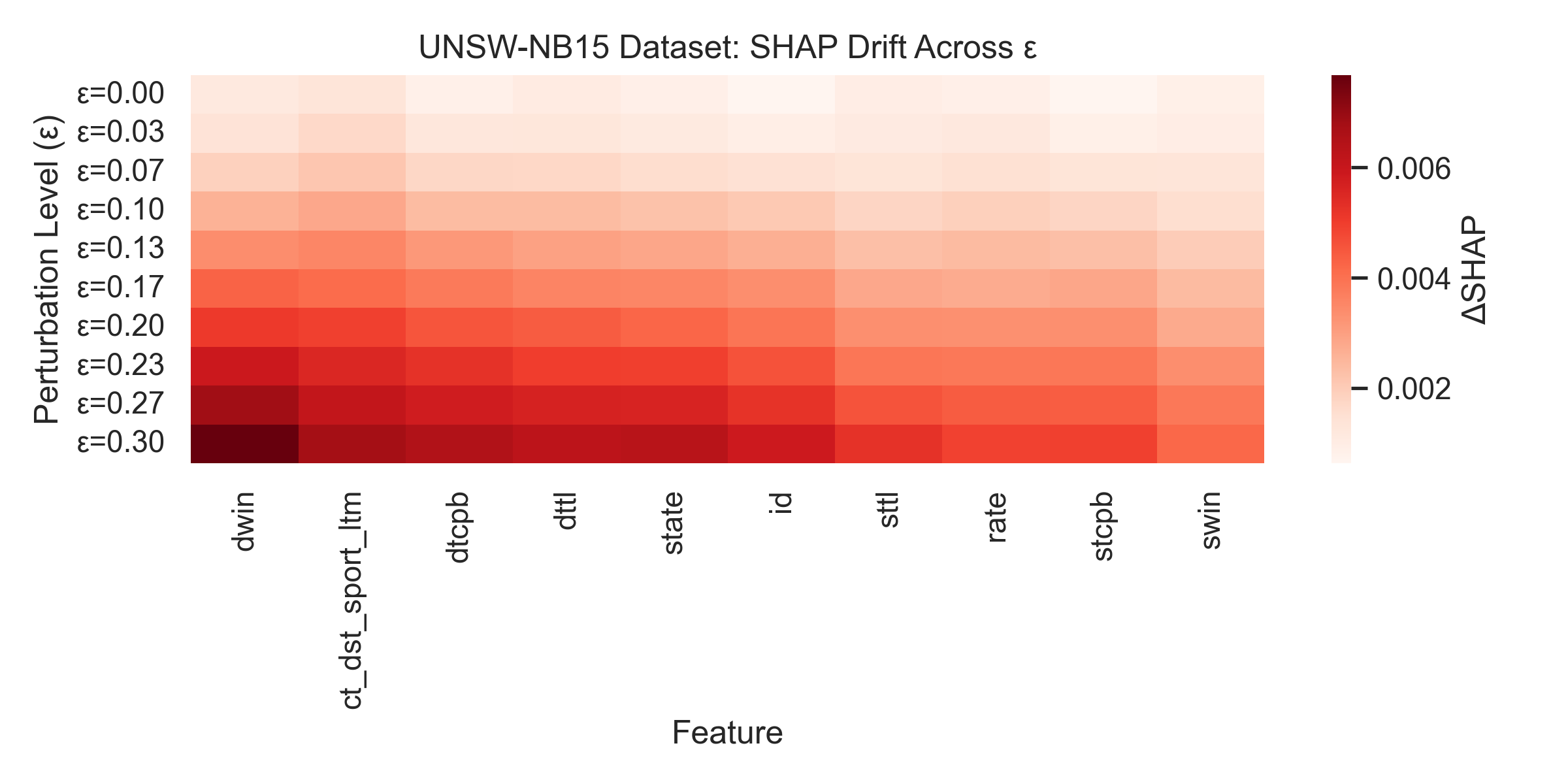}}
\caption{Heatmap of mean SHAP attribution drift $\Delta\phi_i$
for the ten most influential features across $\epsilon\in[0,0.3]$.
Redder regions indicate higher instability in feature importance.}
\label{fig:appendix_heatmap}
\end{figure}

In addition, Figure~\ref{fig:appendix_heatmap} visualizes mean SHAP
attribution drift $\Delta\phi_i$ (Eq.~\ref{eq:shap_drift}) across increasing
perturbation magnitudes.
Each column corresponds to a feature, and color intensity represents the
average change in SHAP value.

These complementary metrics and visualizations further validate that
features with high gradient sensitivity (Eq.~\ref{eq:sensitivity})
also exhibit higher SHAP drift, reinforcing the coupling between robustness
and interpretability degradation observed in Section~V.

\end{document}